\documentclass[times]{article}

\usepackage{moreverb}
\usepackage{amsmath}
\usepackage{graphicx}
\usepackage[utf8]{inputenc}
\usepackage[T1]{fontenc}
\usepackage{hyperref}
\usepackage{lineno}
\usepackage{setspace}
\usepackage{amssymb,amsmath}

\usepackage[ruled,vlined]{algorithm2e}

\newcommand{\R}{\mathbb{R}}
\newcommand{\Z}{\mathbb{Z}}

\newcommand{\C}{\mathcal{C}}
\newcommand{\G}{\mathcal{G}}
\newcommand{\OO}{\mathcal{O}}
\newcommand{\x}{\pmb{x}}

\newcommand{\bb}{\pmb{b}}

\newtheorem{teo}{Theorem}[section]

\begin{document}


\title{A heuristic approach for designing cyclic group codes}

\author{Jo\~ao E. Strapasson and Cristiano Torezzan \\ \small School of Applied Sciences - University of Campinas, SP,Brazil.}

\maketitle

\onehalfspacing

 \begin{abstract}
In this paper we propose a heuristic technique for distributing points on the surface of a unit n-dimensional Euclidean sphere, generated as the orbit of a finite cyclic subgroup of orthogonal matrices, the so called cyclic group codes. Massive numerical experiments were done and many new cyclic group codes have been obtained in several dimensions at various rate. The obtained results assure that the heuristic approach have performance comparable to a brute-force search technique with the advantage of having low complexity, allowing for designing codes with a large number of points in higher dimensions.
\end{abstract}


\maketitle


\vspace{-6pt}

\section{Introduction}
\vspace{-2pt}


Consider a unit vector $\x$ in the $n$-dimensional Euclidean space $\R^n$ and a set of $M$ distinct orthogonal $n \times n$ matrices $G = \{g_1, g_2, \cdots, g_M\}$. The orbit of $\x$ under the action of $ G$ is the set of points $\C= \{ g_i \x, \forall \, g_i \in G \}$. Due to orthogonality, the points in $\C$ belongs to the surface of the unit sphere $S^{n-1} \subset \R^n$. In the case of the set $G$ is a finite cyclic multiplicative subgroup, the orbit $\C$ is called a \textit{cyclic group  code}.


Cyclic group codes belongs to the family of \textit{Slepian} Group Codes, introduced in \cite{Slepian} or, in a more general sense, geometrically uniform codes \cite{for}. Such codes have been widely applied in communication theory, for instance to match signal sets to groups \cite{loe}. In general, point sets on the unit sphere are useful for communicating over a Gaussian channel and are a natural generalization of phase shift keyed signal sets (PSK) to dimensions greater than two \cite{Ingemarsson, Hamkins}.

Improve the design of group codes in higher dimensions may have a considerable impact in error correcting codes and digital communications. In particular, it was proved in \cite{como} that the Shannon capacity of certain important channels, as the AWGN channel with m-PSK modulation, can be achieved using commutative group codes. As we show in Section \ref{sec:results}, cyclic group codes may asymptotically approach the sphere packing bound for commutative group codes, while have a simpler structure. For instance, all the $M$ matrices in a cyclic subgroup of $n \times n$ orthogonal matrices can be generated from only one $\lfloor n/2 \rfloor$-dimensional integer vector (see (\ref{cyclic})  in Section \ref{cycliccodes}).

For any fixed subgroup $G$, the \textbf{initial vector problem} (IVP) is to choose $\x$ which maximizes the Euclidean distance between any two points on the orbit $\C$. This problem has not been solved for general groups, but rather it is known only for special cases such as cyclic groups \cite{biglieri}, commutative groups \cite{torezzan1}, finite reflection groups\cite{Mittelholzer} and permutation codes \cite{eric}. We remark that, for the cyclic and commutative cases, the IVP can be stated as a linear programming problem as reviewed in Section \ref{IVP}.

An important quota in the theory of spherical codes is the maximum number of points displayed on the surface of a unit sphere with distance not smaller than a certain $d$. In \cite{siq} it is derived the bound (\ref{limcom})  on the maximal number $M$ of points in a $(2k)$-dimensional commutative group code as a function of the minimal distance $d$ and the maximal center density of a lattice packing in $k$-dimension. In this case, 
\begin{equation}
\displaystyle M \leq \frac{\pi^k
\sqrt{\Pi_{i=1}^{k}(x_{2i-1}^2+x_{2i}^2)}
\Lambda_{k}}{(\arcsin{\frac{d}{4}})^k} \leq \frac{\pi^k \Lambda_{k}}
{(\arcsin{\frac{d}{4}})^k k^{k/2}} ,
\label{limcom}
\end{equation}
where $\Lambda_{k}$ is the maximal center density of a lattice packing in $\R^{k}$ and $\x$ $=(x_{1},x_{2}, \ldots, x_{2n})$ is the initial vector.

Despite the relevant literature on group codes (indeed spherical codes in general\footnote{A thorough survey of spherical codes can be found in \cite{eric} and explicit constructions of spherical codes were proposed in \cite{hamk1, hamk2, torezzan}. Lists of good spherical codes and also packings, coverings and designs can be found online at \cite{sloane}.}), explicit construction of cyclic group codes for a large number of points is still a challenge problem, particularly in high dimensions. It is worth to remark that the classical \textit{Simplex} code is a cyclic group code with $n+1$ points in $n$-dimension and the \textit{Biorthogonal} code is also a cyclic group code with $M=2n$ points in odd dimension (\textit{Biorthogonal} is not cyclic in even dimension). A more detailed discussion can be found in Chapter 8 of \cite{eric}.

In \cite{torezzan1} it is presented a method for finding an optimum $(2k)$-dimensional commutative group code of a given order $M$, which is based on the structure of lattices related to these codes. Although that method provides a significant reduction in the number of non-isometric cases to be analyzed, it still demands a brute-force search on about 
\begin{equation}
\frac{M^n}{2^n \varphi(M)}
\label{numcases}
\end{equation}
cases, where $\varphi$ stands for the \textit{Euler phi} function of $M$. 

It is easy to see that, when $M$ and $n$ increase, the number (\ref{numcases}) become computationally prohibitive. For instance, for $(2k,M) = (16,1024)$, which corresponds to a commutative group code in $S^{15}\subset \R^{16}$, with $M=1024$ points, the approach proposed in \cite{torezzan1} to find the optimal code demands a full search in a set with about $2^{63}$ elements. In addition, it also means solving $2^{63}$ linear programming problems to find the best initial vector for each case.

In this paper, we present a heuristic technique inspired on \cite{torezzan1}, for designing $(2k)$-dimensional cyclic group codes for a given number of points $M$, avoiding the brute-force search and hence allowing for find codes with a large number of points in high dimensions. Using this technique we were able to design new cyclic group codes for $M$ up to $2^{19}=524288$ points in dimensions up to $n=48$, with good performance (Figure \ref{fig:resulall}).

Although the heuristic can be adjusted for design commutative group codes, we restrict our attention in the cyclic case because these codes have a simpler and stronger structure and, moreover, the codes founded with our heuristic suggests a conjecture that the cyclic family can asymptotically approach the sphere packing bound for commutative group codes (Figure \ref{Fig_app_bound}).

The rest of the paper is organized as follows: In Section \ref{cycliccodes} we review fundamental concepts on cyclic group codes and set our notation. In Section \ref{heuristic} we describe our heuristic method and present a pseudo-code algorithm for that. In Section \ref{sec:results} we show various cyclic group codes designed using the heuristic approach and some performance comparisons. Finally, in Section \ref{sec:conclusion} we summarize and conclude the paper.

\section{Cyclic group codes}
\label{cycliccodes}

Let $\OO_n$ be the multiplicative group of or\-tho\-go\-nal matrices $n \times n$ and $\G_n(M)$ be the set of all order $M$ cyclic subgroups in $\OO_n$.

A \textit{cyclic group code} $\C$ on the surface of a $n$- dimensional euclidean sphere is a set of $M$ vectors generated as the orbit of an initial vector $\x \in S^{n-1} \subset \R^n$ by a given $G \in \G_n(M)$, i.e. 

$$\C := G \x = \left\{g \x, g \in G \right\}.$$


The \textit{minimum distance} in $\C$ is defined as:

$$
\displaystyle d  := \min_{
\scriptsize
\begin{array}{c}
\x, \pmb{y} \in \C \\ 
\x \neq \pmb{y}
\end{array} } ||\x-\pmb{y}|| = \min_{\scriptsize
\begin{array}{c}
g_i \in G \\ 
g_i \neq I_n
\end{array}} ||g_i \x - \x||,
$$
where $\displaystyle ||.||$ denote the standard Euclidean norm and $I_n$ is the identity matrix of order $n$.

We use $\C(M,n,d)$ to denote a code $\C$ in $\R^n$ with $M$ points and minimum distance equal to $d$. 

The minimum distance of a cyclic group code $\C$, depends on the initial vector $\x$ and also on the orthogonal matrices on $G$ (in other words, the distance also depends on the real representation of the subgroup $G$). For every finite cyclic subgroup $G$ it is possible to find a generator matrix $\bar{g}$, such that $G = \lbrace (\bar{g})^j \rbrace_{j=0}^{M-1}$, where $(\bar{g})^0=I_n$ is the identity matrix in $\R^n$. Thus, $\C = \lbrace (\bar{g})^j \x \rbrace_{j=0}^{M-1}$ and therefore $$\displaystyle d = \min_{\scriptsize
\begin{array}{c}
j \neq 0\\ 
\end{array}} || (\bar{g})^j \x - \x||.$$

For every commutative group code of orthogonal matrices there is a well known real-irreducible representation, which can be stated as follows:
\begin{teo}(\cite{gan} Theorem 12.1) Every commutative group $G \in \G_n(M)$ can be carried by the same real orthogonal transformation $q$ into a pseudo-diagonal form:
\label{teogan}
{

$$
	qg_{_i}q^{t}=[R_i(1),\ldots,R_i(k),\mu(i)_{2k+1},\ldots,\mu(i)_{n}]_{n\times n},
$$

\begin{equation}
\label{pseudo_diagonal}
\mbox{where  }R_{i}(b_j)=\left[
\begin{array}
[c]{cc}%
\cos(\frac{2\pi b_{ij}}{M}) & -\sin(\frac{2\pi b_{ij}}{M})\\
\sin(\frac{2\pi b_{ij}}{M}) & \cos(\frac{2\pi b_{ij}}{M})
\end{array}
\right],
\end{equation}

$$b_{ij} \in Z, \ \ 0 \leqslant b_{ij} \leqslant M \mbox{ and }\mu(i)_{l}=\pm 1  \mbox{, } l = 2k+1, \hdots, n, \, j = 1, \hdots, k, \forall g_i \in G.$$
}
\end{teo}

This result holds, in particular, for the generator $\bar{g}$, which means that every cyclic group code in even dimensions is equivalent to one in which the generator matrix $\bar{g}$ has a quasidiagonal form 
\begin{equation}
\bar{g} = [R(b_1),\ldots,R(b_{n/2})]_{n\times n}, \mbox{ where } R(b_j) = \left[\begin{smallmatrix} \cos\big(\frac{2\pi b_{j}}{M}\big) & -\sin\big(\frac{2\pi b_{j}}{M}\big)\\ \sin\big(\frac{2\pi b_{j}}{M}\big) & \cos\big(\frac{2\pi b_{j}}{M}\big) \end{smallmatrix}\right].
\label{cyclic}
\end{equation}

Note that each $2 \times 2$ block $R(b_j)$ in (\ref{cyclic}) is a rotation matrix by an angle of $\left(2 \pi b_{j}/M \right)$. Thus the generator matrix $\bar{g}$ of any cyclic group $G \in \OO_{n=2k}$ is defined by a vector \linebreak $\bb=\big(b_1,b_2,\cdots, b_k \big)$ with $0 < b_i \leq M$ and $\gcd(b_1,b_2,\cdots, b_k,M)=1$ that represent the rotation blocks. In addition, as pointed out in \cite{torezzan1} up to conjugacy we can consider $0 < b_j \leq M/2$.

If $n$ is odd, then $$
\bar{g} = [R(b_1),\ldots,R(b_{\lfloor k \rfloor }), \mu]_{n\times n},
$$
where $\mu = 1$ if $M$ is even and $\mu = -1$ if $M$ is odd \cite{biglieri}. Thus, the generator $\bar{g}$ may be always defined by the vector $\bb$.

\subsection{The Initial Vector Problem for cyclic group codes}
\label{IVP}
Using the decomposition (\ref{cyclic}) it is possible to state the initial vector problem (IVP) for commutative group codes (hence for cyclic) as a linear programming problem as follows\cite{biglieri, torezzan1}:  

For a given $G \in \OO_n$ the initial vector problem is to find $\x \in S^{n-1}$ such that:
$$
{\max_{\x\in S^{n-1}}} \left( {\min_{i \neq 0}} ||g_i \x - \x||^2 \right)
$$

From Theorem \ref{teogan} and Equation (\ref{pseudo_diagonal}) we can define
$$
f_i(\x) := ||g_i \x - \x||^2 = 4\sum_{j=1}^{q}(x_{2j-1}^2+x_{2j}^2)\sin^2{\frac{\pi b_{i,j}}{M}}+4 \sum_{j=q+1}^{n}(1-\mu(i)_j)^2x_j^2, \, i = 1, \ldots, M.
$$
Let
$$y_j= \left\{ \begin{array}{cl} (x_{2j-1}^2+x_{2j}^2) & \mbox{, if } j=1,\ldots,k \\ 
x_{j+k}^2 & \mbox{, if }j=k+1,\ldots,n-k \end{array}\right. ,$$ 
we have that
$$
f_i(y) = 4 \sum_{j=1}^{k}y_j \left( \sin^2{\frac{\pi b_{i,j}}{M}} \right) + 4 \sum_{j=k+1}^{n-k} y_j \left( 1-\mu(i)_{j+k}\right)^2, 
$$
and, therefore
$$
{\max_{\x \in S^{n-1}}} \left( {\min} ||g_i \x - \x||^2 \right) = {\max_{y\in S^{(n-k-1)}}} \left( {\min} f_i(\pmb{y}) \right), \mbox{ where }, \pmb{y} = (y_1, \cdots, y_k).
$$

This max-min problem can be stated as a linear programming problem as in \cite{torezzan1}:
$$\min{c^t w}$$
$$
\mbox{s.t.}
\left\{
\begin{array}{c}
 Aw \leq b \\
 w\geq 0
\end{array}
\right.
$$
where:
\begin{eqnarray*}
c^t = & (-1,0,\dots,0) \in \R^{n-k+1}\\
b^t = & (1,-1,0,\dots,0) \in \R^{\left\lfloor \frac{M}{2}\right \rfloor +2} \\
A  = & \left(
\begin{array}{cccc}
 0 & 1 & \ldots  & 1 \\
 0 & -1 & \ldots  & -1 \\
 1 & m_{1,1} & \ldots  & m_{n-q,1} \\
 \vdots  & \vdots  & \ddots & \vdots  \\
 1 & m_{1,\left \lfloor \frac{M}{2}\right\rfloor} & \ldots  & m_{n-k,\left\lfloor \frac{M}{2}\right\rfloor}
\end{array}
\right)\\
m_{i,j} = & \mbox{ is the }  (j-\mbox{th})\mbox{ coeficient of} f_i.
\end{eqnarray*}

Remark: since this LP-problem has $(n-k+1)$ variables and $\lfloor M/2 \rfloor$ constraints, when $M>>n$ it is worth to solve this by using the dual-simplex method. This is the method we have used for running our numerical experiments presented in Section \ref{sec:results}.

\subsection{On optimal cyclic group codes}

From the previous discussion we can conclude that for a given number of points $M$ it is possible to run a brute-force search method  for finding an $n$-dimensional cyclic code by analyzing all possible vectors $\bb$ and solve the correspondent IVP for each one of them. In fact this approach was used in \cite{biglieri} for find some optimal cyclic group codes for small number of points $M$ and improved for commutative group codes in \cite{torezzan1}.

The main weakness of the brute-force approach is that it leads to check about $\displaystyle \genfrac{(}{)}{0pt}{}{M/2}{n/2}$ cases and, for each one of them, solving the IVP to maximize the minimum distance. Indeed, it has been shown in \cite{torezzan1} that it is possible to reduce the number of cases to less than (\ref{numcases}), by considering isometries arising from lattices. Using this reduction, it was presented in \cite{torezzan1} several optimum commutative group codes in dimensions $4$ and $6$ for $M \leq 1000$. Neverthless, as pointed out in the introduction, the number (\ref{numcases}) may become computationally prohibitive when $M$ increases.

In the next section we describe a heuristic approach that allowed us to design codes for $n$ up to $48$ with $M\leq 2^{19}$ points. As we can see in Table \ref{comparaotimo}, the codes found by the proposed heuristic have comparable performance with some optimal codes known and are possibly the best published lower bounds for many values of $(M,n)$.

\section{A heuristic approach for designing cyclic group codes}
\label{heuristic}

We start with the same approach as in \cite{torezzan1}, i.e. by exploring the connection between cyclic group codes and lattices. Formally, let $\C$ be a cyclic group code in $\R^{n=2k}$, generated by a group $G \in \OO_n$,
 $G = \lbrace (\bar{g})^j \rbrace_{j=0}^{M-1}$, where $\bar{g}$ is defined as in (\ref{cyclic}) by an integer vector $\bb = (b_1, \cdots, b_k)$.

The associated lattice $\Lambda_G$ is defined by
$$
\Lambda_G := \left\lbrace (b_1, \hdots, b_k) \in \Z^k: [R(b_1), \hdots, R(b_k)] \in G \right\rbrace,
$$
where $R(b_j)$ denotes the rotation in $\R^2$ by an angle of $2 \pi b_j/M$ and $[R(b_1), \hdots, R(b_k)]$ denotes a pseudo-diagonal matrix, according to  (\ref{cyclic}).

As pointed out in \cite{torezzan1} $\Lambda_G$ contains $M \Z^k:= \left\lbrace M (z_1,z_2,\hdots,z_k), z_i \in \Z \right\rbrace$ as a sub-lattice. Thus, inside the hyperbox $[0,M)^k$ there are exactly $M$ points of $\Lambda_G$, which correspond to representatives of the elements of $G$, i.e., 
$$[0,M)^n \supset \left\lbrace (b_{1},b_{2},\hdots, b_{k})\mod M, \ \ i=1,2,\hdots,M \right\rbrace.$$


If $\x=(x_1, 0,x_2, 0, \dots , x_{k}, 0)$ is an initial vector for the code $\C$, we can also define a lattice $\Lambda_G(\x)$ by
$$
\Lambda_G(\x) := \left\lbrace \left[\begin{smallmatrix} \frac{2 \pi \x_1}{M} & & & \\ & \frac{2 \pi \x_2}{M} & & \\ & & \ddots & \\ & & & \frac{2 \pi  x_{k}}{M} \end{smallmatrix}\right] b: b \in \Lambda_G \right\rbrace.
$$

Under this assumption the code $\C$ is the image $\psi_{\x}(\Lambda_G(\x)) \subset S^{2k-1}$, where
\begin{equation}
\label{lambdaGG}
\psi_{\x}(y)=\displaystyle \left(x_{1}\cos\left(\frac{y_{1}}{x_{1}}\right),x_1\sin\left(\frac{y_{1}}{x_{1}}\right),\ldots,x_{k}\cos\left(\frac{y_{k}}{x_{k}}\right),x_k\sin\left(\frac{y_{k}}{x_{k}}\right)\right)
\end{equation}
is the standard parametrization of the torus with radii $x_i, \, i = 1,\ldots, k$ \cite{siq}.

\begin{figure}[h!]
\centering
		\includegraphics[width=6cm]{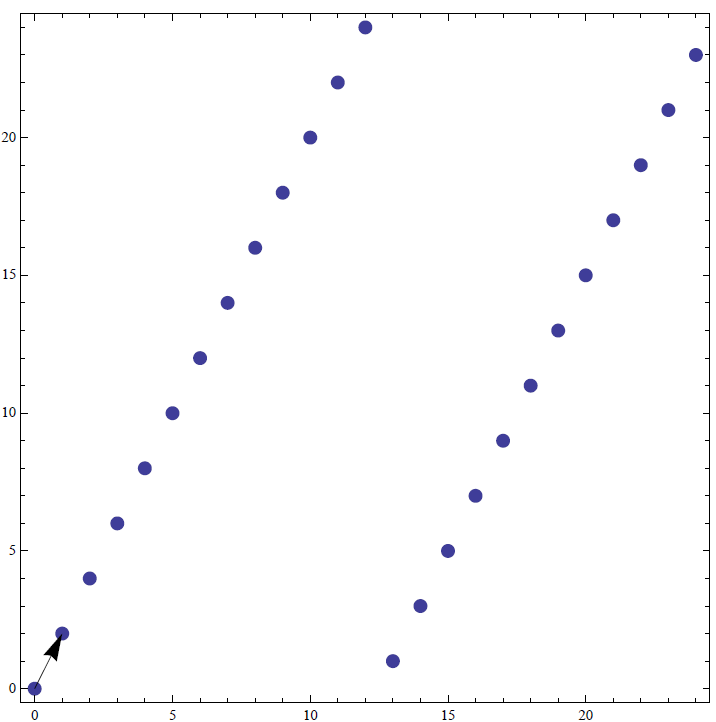}
		\includegraphics[width=6cm]{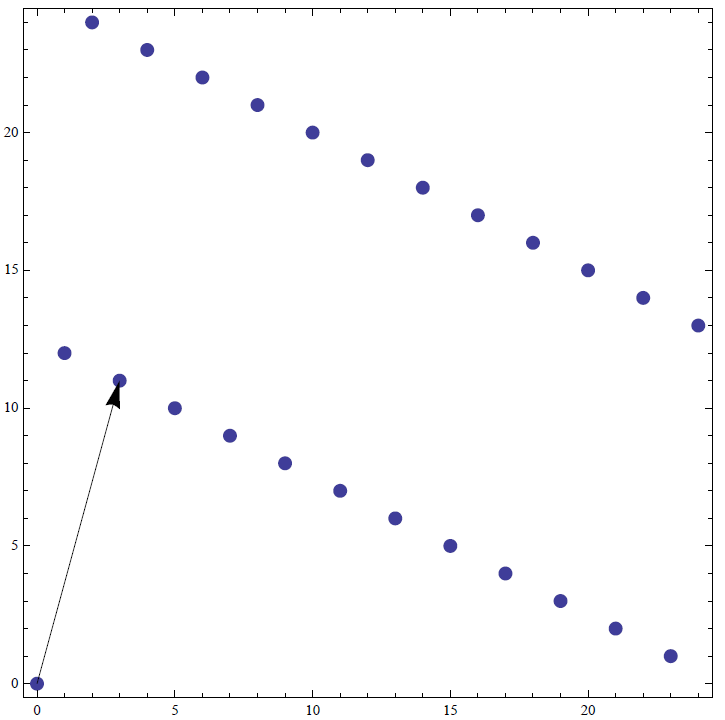}
	\caption{The associated $2$-dimensional lattices $\Lambda_1$ and $\Lambda_2$ of two isometric $4$-dimensional cyclic group codes $\C_1$ and $\C_2$ defined by the vectors $\bb_1 = (1,2)$ and $\bb=(3,11)$, respectively.}
		\label{Fig:IsometryCodes}
\end{figure}

As an example, let us consider two $4$-dimensional cyclic group codes $\C_1$ and $\C_2$, both with $M=25$ points, defined by vectors $\bb_1 = (1,2)$ and $\bb=(3,11)$, respectively. These two particular codes are isometrics and so their associated lattices $\Lambda_1$ and $\Lambda_2$ are isomorphic, as illustrated in \ref{Fig:IsometryCodes}.

This association between a $2k$-dimensional cyclic group codes and a $k$-dimensional lattices is behind our heuristic method, as we describe in the next Section.

\subsection{The heuristic}

The key idea: 

When we analyze the lattice $\Lambda_1$, plotted on the left side of Figure \ref{Fig:IsometryCodes}, which correspond to the group defined by the vector $\bb=(1,2)$, we may conjecture that the minimal distance in the $4$-dimensional group code $\C_1$, is given by $||\psi_{\x}(0,0)- \psi_{\x}(1,2)||$. Thus, if we want to find codes with large minimal distances, we should  discard candidates such that $||\bb||$ is to small. On the other hand, we must hold our greed, because if $||\bb||$ is bigger than a certain threshold, then the minimal distance of the group code will be defined by other pair of points, rather then $||\psi_{\x}(0,0)- \psi_{\x}(1,2)||$, as occurs, for instance,  in the lattice $\Lambda_2$ (right hand side of Figure \ref{Fig:IsometryCodes}).

In what follow we derive a fair threshold for $||\bb||$ in order to get good cyclic group codes, which is the basis for our heuristic approach.

Since vector $\pmb{b}=(b_1, \dots, b_k)$  defines a generator $\bar{g}$ of a cyclic subgroup $G$, if and only if, $\gcd(b_1, \dots, b_k,M)=1$, we can trivially guarantee this condition by choosing $b_1=1$ and, of course, reducing the set of candidates (eventually loosing optimality). So in the heuristic we assume that $b_1=1$ and it is the first reduction comparing to a force-brute approach.

The main reduction in the set of candidates comes from writing the inequality (\ref{limcom}) in terms of $d$, 
\begin{equation}
\label{disttarg}
d \leq 4 \sin{ \left( \frac{\pi^k \Lambda_k}{M k^{k/2}} \right)^{1/k}},
\end{equation}
and get a target (upper bounded) distance $\check{d}:=4 \sin{ \left( \frac{\pi^k \Lambda_k}{M k^{k/2}} \right)^{1/k}}$ in the cyclic group code $\C(M,n)$. 

We know that $\Lambda_k = \dfrac{||\pmb{v}_{min}||^k}{2^k \text{vol}(\Lambda)}$, where $||\pmb{v}_{min}||$ and vol$(\Lambda)$ are the minimal norm and the volume of the densest $k-$dimensional lattice, respectively. Thus, for lattices associated to cyclic groups we can approach $||\pmb{v_{min}}|| = ||\bb||$ and consider vol$(\Lambda) = M^{k-1}$. 


Thus, we may restrict our search on candidates $\bb$ such that $b_1 =1$ and
\begin{equation}
\label{cand_b}
||\bb\| \simeq \frac{2 M  \sqrt{k}}{\pi} \arcsin(\check{d}/4)
\end{equation}

\begin{figure}[h!]
\centering
		\includegraphics[width=7cm]{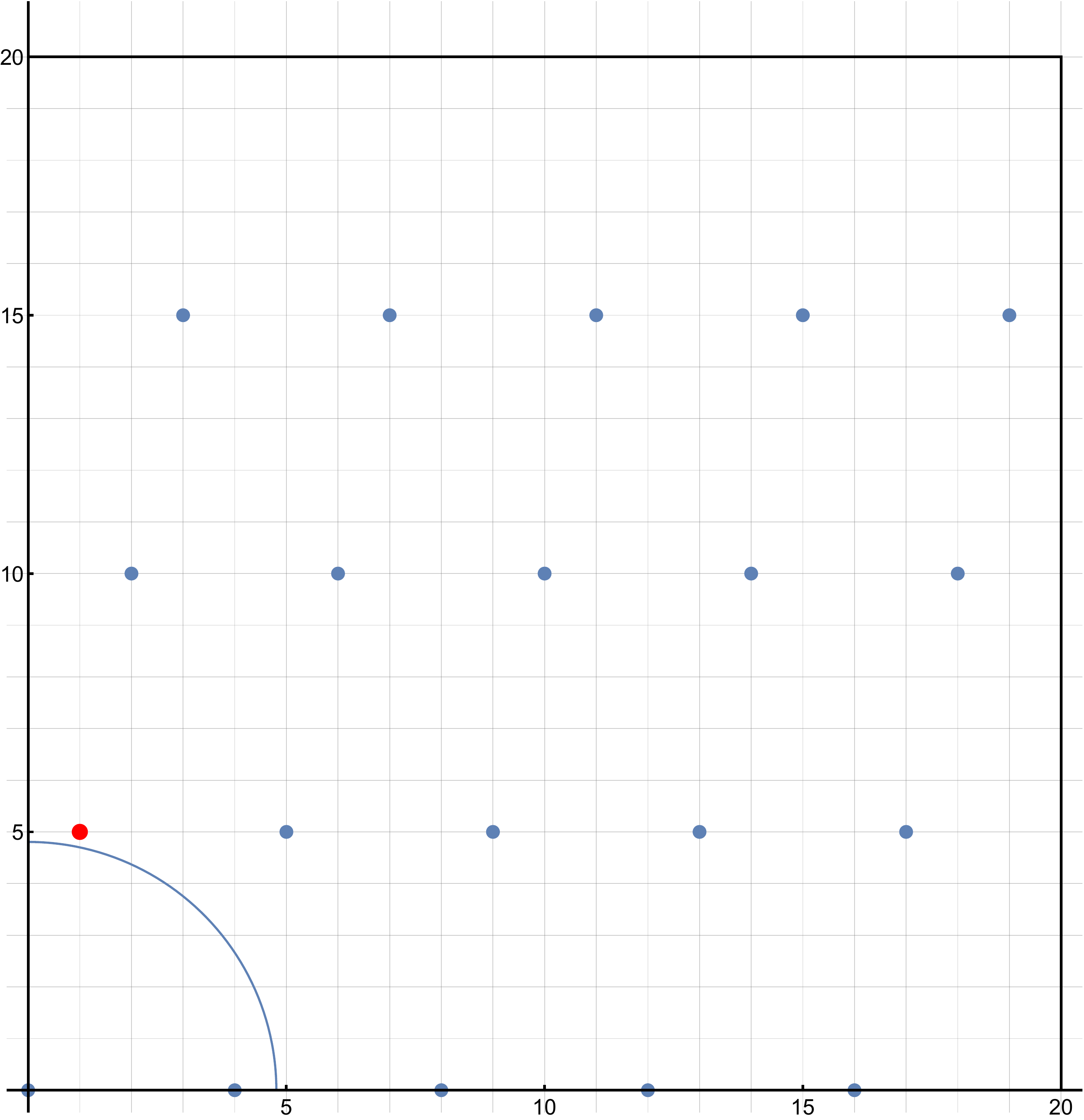}
	\caption{The associated lattices in $\R^2$ of a $4$-dimensional cyclic group codes with $M=20$ points defined by the vector $\bb = (1,5)$ and an arc-circle of radius $\check{d}$.}
		\label{Fig:cic2d}
\end{figure}

In Figure \ref{Fig:cic2d} we show an example of the choice of candidates according this heuristic for finding a $4$-dimensional cyclic group code with $M=20$ points. The circle (arc) has radius $\check{d} = \simeq 4.805$. In this case, the only vector to be tested is $\bb = (1,5)$ which has $||\bb||\simeq 5.1 \simeq \check{d}$. This vector corresponds to a cyclic group code with minimum distance equal to $1.19$ while the best commutative group code for $(M,n) = (20, 4)$ found in \cite{torezzan1} has distance equal to $1.24$ and the bound (\ref{disttarg}) is equal to $1.465$.

\begin{algorithm}
\SetKwData{Left}{left}\SetKwData{This}{this}\SetKwData{Up}{up}
\SetKwFunction{Union}{Union}\SetKwFunction{FindCompress}{FindCompress}
\SetKwInOut{Input}{inputs}\SetKwInOut{Output}{outputs}
\Input{The number of points $M$; The dimension $n=2k$; The maximal center density of a lattice packing in $\R^k$ $\Lambda_k$;  A parameter $Q$ to define the number of cases to be tested}
\Output{A vector $\bb$, corresponding to the generator of a cyclic group code $\C(M,n)$; The optimum initial vector $\x$ and minimum distance $d$.}
\BlankLine
\Begin{
initialization\;
 $d_{best} \leftarrow 0$\;
 $\pmb{v}_{best} \leftarrow \emptyset$\;
 $\x \leftarrow \emptyset$\;
$\displaystyle \rm{\check{d}} \leftarrow 4 \sin\left(\frac{\pi^k \Lambda_k}{M\, k^{k/2}}\right)^{1/k} $\;
 $\displaystyle  W \leftarrow \frac{2 M  \sqrt{k}}{\pi} \arcsin(\check{d}/4)$\; 
 \While{$C>0$}{
1) Chose a random vector $\pmb{v}=(v_1, \dots, v_k)$, such that: $\displaystyle v_1=1, v_i\leqslant v_{i+1} \leqslant \sqrt{\dfrac{W^2-v_1^2- \cdots -v_i^2}{k+1-i}}$ and $\displaystyle  v_k= \Bigl\lfloor \sqrt{W^2-v_1^2- \cdots -v_{k-1}^2} \, \, \Bigr\rceil  $\;
 2) Solve the LP corresponding to the IVP to get $\x$ and the minimum distance $d_{min}$\;
  \If{$d_{min}>d_{best}$}{
  $d_{best} \leftarrow d_{min}$ \;
  $\pmb{v}_{best} \leftarrow \pmb{v}$\;
  $\x_{best} \leftarrow \x$\;
  }
 $C \leftarrow C-1$\;
 }
 $d \leftarrow d_{best}$\;
 $\bb \leftarrow \pmb{v}_{best}$\;
  $\x \leftarrow \x_{best}$
}
\caption{Heuristic method for design cyclic group codes}
\label{algo_disjdecomp}
\end{algorithm}

We remark that, in the $2$-dimension, the only choice we have is on coordinate $b_2$, but when the dimension increases, it become more flexible, allowing testing more candidates $\bb$ and hence increasing the chance of obtaining good codes.

Algorithm 1 summarize a pseudo-code for the heuristic  and some results are described in the next Section.

\section{Results}
\label{sec:results}
In this section we present numerical results obtained with the heuristic described in the previous section. Various codes were found in dimensions up to $48$ for $M$ up to $2^{19}$.

\begin{figure}[h!]
\centering
	\includegraphics[scale=.5]{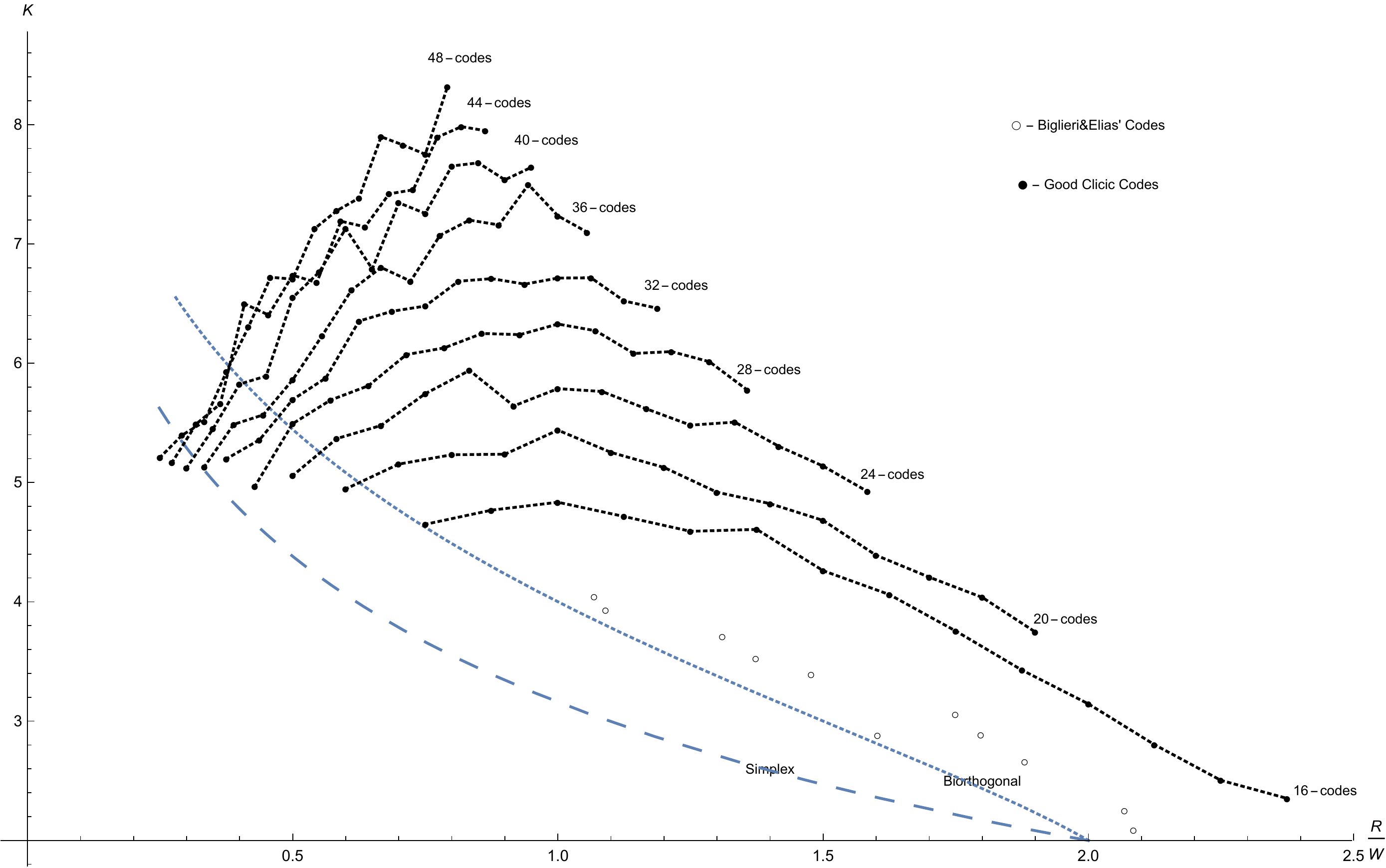}
\caption{Efficiency comparison for commutative group codes, for 1000 cases analysing.}
\label{fig:resulall}
\end{figure}

In Figure \ref{fig:resulall} we present a compilation of several cyclic group codes found in this paper. The results have been plotted, as in \cite{biglieri}, in a Jabobs' efficiency chart for comparison. In this chart, every code $\C(M,n)$ with minimal distance $d$ and minimal angle separation $\rho = 2 \arcsin(d/2)$ is represented by a point $(R,K)$, where \linebreak $K = (1-\rho) \log_2 M$ and $R = \frac{2}{n}\log_2 M$. This charts exhibit codes in dimensions $8, 10, \cdots, 24$ for $M =  2^6, 2^7, \cdots, 2^{19}$ points. The dotted and dashed lines correspond to the performance of biorthogonal and simplex codes, respectively.

As we can see, most of the codes presented in \ref{fig:resulall} outperform the classical Simplex and Biorthogonal codes, and also the codes showed in \cite{biglieri}. Therefore, using the heuristic we were able to design cyclic group codes in high dimension with a large number of points that really have good performance. We believe the codes presented in this paper might be the best known lower bounds for minimal distances of cyclic group codes in several dimensions published up to now.

\begin{table}[h!]
\centering
$\begin{array}{|c|c|c|c|c|}
\hline
 \text{M} & Q = 512 & Q =1024  & Q =1536  & Q =2048 \\
 \hline
 \hline
 64 & 1.27269 & 1.28386 & 1.2859 & 1.29514 \\
 128 & 1.21733 & 1.21962 & 1.23481 & 1.23481 \\
 256 & 1.15941 & 1.16528 & 1.16905 & 1.23721 \\
 512 & 1.13819 & 1.19334 & 1.19334 & 1.19334 \\
 1024 & 1.11273 & 1.12457 & 1.12457 & 1.12457 \\
 2048 & 1.09191 & 1.09191 & 1.09191 & 1.09191 \\
 4096 & 1.06506 & 1.06506 & 1.07339 & 1.07339 \\
 8192 & 0.99974 & 1.02577 & 1.02577 & 1.03603 \\
 16384 & 1.02518 & 1.03103 & 1.03103 & 1.03103 \\
 32768 & 0.98699 & 1.01643 & 1.01643 & 1.01643 \\
 65536 & 0.99445 & 0.99445 & 0.99445 & 0.99445 \\
 131072 & 0.94748 & 0.94748 & 0.96942 & 0.96942 \\
 262144 & 0.95448 & 0.95448 & 0.95565 & 0.95966 \\
 524288 & 0.91805 & 0.95606 & 0.95606 & 0.95606 \\
 \hline
\end{array}$
\caption{Heuristic performance depending on the number of candidates $Q$ tested for each case. Examples for designing $48$-dimensional codes whit $M = 2^6, 2^7, \cdots, 2^{19}$.}
\end{table}

\begin{table}[h!]
$\begin{array}{|c|c|c|c|}
\hline
 \text{M} & \text{Bound} & \text{Optimum} & \text{Heuristc} \\
 \hline
 \hline
 10 & 1.474 & 1.224 & 1.224 \\
 20 & 1.054 & 0.959 & 0.917 \\
 30 & 0.864 & 0.831 & 0.769 \\
 40 & 0.750 & 0.714 & 0.707 \\
 50 & 0.672 & 0.628 & 0.609 \\
 100 & 0.476 & 0.468 & 0.433 \\
 200 & 0.337 & 0.330 & 0.317 \\
 300 & 0.275 & 0.273 & 0.259 \\
 400 & 0.238 & 0.237 & 0.221 \\
 500 & 0.213 & 0.211 & 0.200 \\
 600 & 0.194 & 0.193 & 0.180 \\
 700 & 0.180 & 0.180 & 0.167 \\
 800 & 0.168 & 0.168 & 0.162 \\
 900 & 0.159 & 0.158 & 0.148 \\
 1000 & 0.150 & 0.149 & 0.146 \\
 \hline
\end{array}$
\hfill
$
\begin{array}{|c|c|c|c|}
\hline
 \text{M} & \text{Bound} & \text{Optimum} & \text{Heuristc} \\
 \hline
 \hline
 10 & 1.820 & 1.414 & 1.345 \\
 20 & 1.465 & 1.240 & 1.190 \\
 30 & 1.287 & 1.133 & 1.056 \\
 40 & 1.173 & 1.044 & 1.007 \\
 50 & 1.091 & 0.976 & 0.946 \\
 100 & 0.870 & 0.804 & 0.786 \\
 200 & 0.692 & 0.673 & 0.633 \\
 300 & 0.605 & 0.585 & 0.568 \\
 400 & 0.550 & 0.540 & 0.525 \\
 500 & 0.511 & 0.504 & 0.479 \\
 600 & 0.481 & 0.472 & 0.458 \\
 700 & 0.457 & 0.445 & 0.439 \\
 800 & 0.437 & 0.427 & 0.415 \\
 900 & 0.420 & 0.413 & 0.403 \\
 1000 & 0.406 & 0.397 & 0.394 \\
 \hline
\end{array}$
\caption{Comparison between codes found by the heuristic for several values of $(M,n)$ with respective optimum cases and bound.}
\label{comparaotimo}
\end{table}

\begin{figure}[h!]
\centering
		\includegraphics[scale = 0.5]{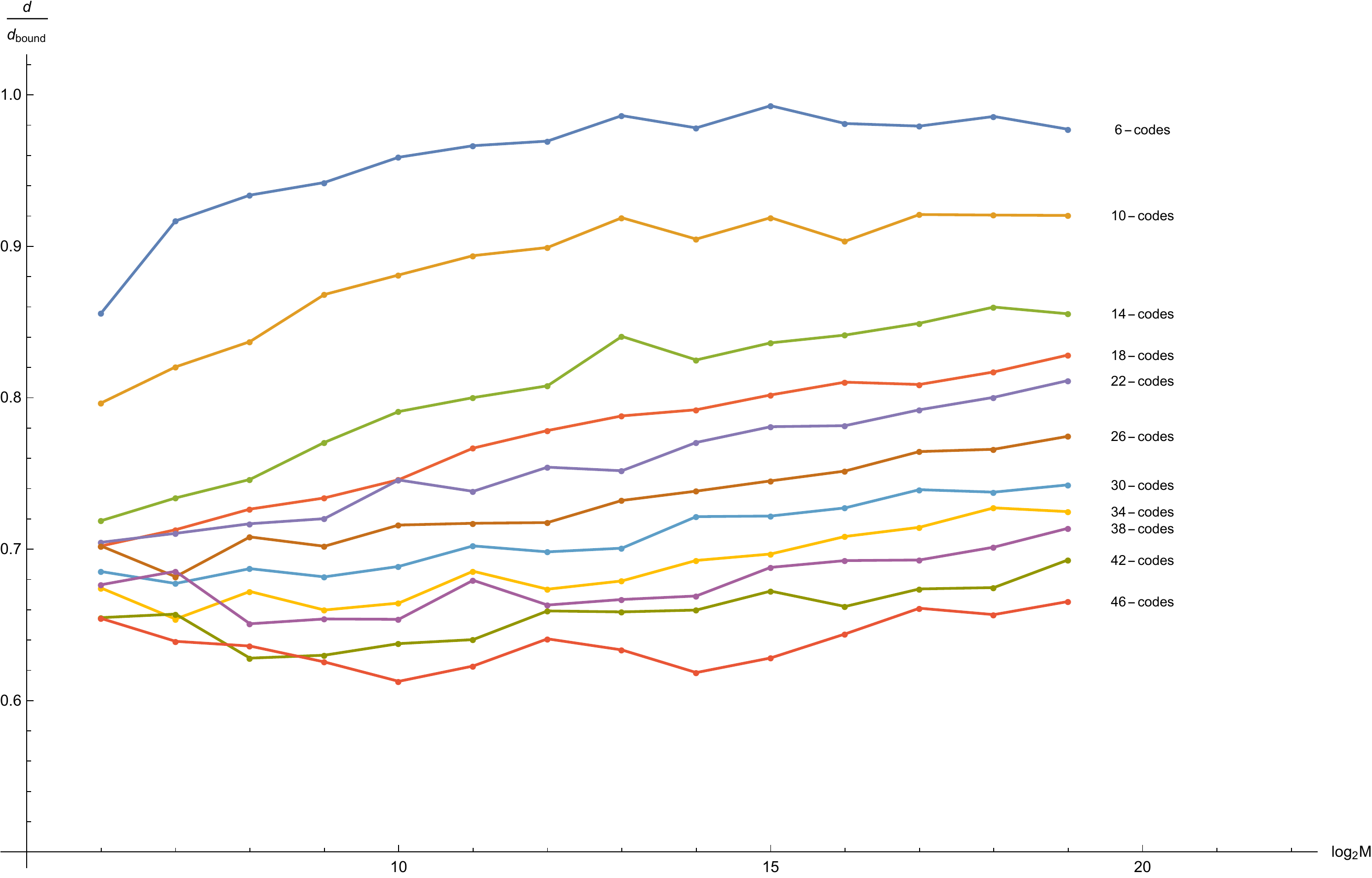}
	\caption{Comparison between the minimal distance of cyclic group codes found by using the heuristic and the sphere packing bound for commutative group codes.}
		\label{Fig_app_bound}
\end{figure}

In Figure \ref{Fig_app_bound} we compare the ration between the minimal distance $d$ of various cyclic group codes found by using the heuristic and the target distance $\check{d}=d_{bound}$ obtained from the sphere packing bound (\ref{disttarg}). In the $x$-axis we plotted $\log_2(M)$ and the y-axis show $\frac{d}{\check{d}}$. As we can see, when $M$ increases the ratio approaches to $1$, but the convergence is slower as the higher is the dimension.

\section{Conclusions}
\label{sec:conclusion}

We presented a heuristic technique for designing $n=2k$-dimensional cyclic group codes for given number of points $M$. The heuristic explores the relation between such codes and $k$-dimensional lattices. Numerical experiments were done and many new cyclic group codes have been obtained in several dimensions at various rate. The obtained results assure that the heuristic approach have performance comparable to a brute-force technique with the advantage of having low complexity, which allows for designing codes with large number of points in higher dimensions. The results also suggest that the sphere packing bound for commutative group codes can be asymptotically approached by cyclic group codes.

\end{document}